\documentstyle[prd,aps,preprint,tighten,epsfig]{revtex}

\begin{document}

\draft

\title{Possible Deviation from the Tri-bimaximal Neutrino Mixing \\
in a Seesaw Model}
\author{{\bf Sin Kyu Kang}
\thanks{E-mail: skkang@phya.snu.ac.kr}}
\address{School of Physics, Seoul National University,
       Seoul 151-734, Korea}
\author{{\bf Zhi-zhong Xing}
\thanks{E-mail: xingzz@mail.ihep.ac.cn} ~ and ~ {\bf Shun Zhou}
\thanks{E-mail: zhoush@mail.ihep.ac.cn}}
\address{CCAST (World Laboratory), P.O. Box 8730, Beijing 100080,
China \\
and Institute of High Energy Physics, Chinese Academy of Sciences,
\\
P.O. Box 918, Beijing 100049, China
\footnote{Mailing address}}
\maketitle

\begin{abstract}
We propose a simple but suggestive seesaw model with two
phenomenological conjectures: three heavy (right-handed) Majorana
neutrinos are degenerate in mass in the symmetry limit and three
light Majorana neutrinos have the tri-bimaximal mixing pattern
$V^{}_0$. We show that a small mass splitting between the first
generation and the other two generations of heavy Majorana
neutrinos is responsible for the deviation of the solar neutrino
mixing angle $\theta^{}_{12}$ from its initial value $35.3^\circ$
given by $V^{}_0$, and the slight breaking of the mass degeneracy
between the second and third generations of heavy Majorana
neutrinos results in a small mixing angle $\theta^{}_{13}$ and a
tiny departure of the atmospheric neutrino mixing angle
$\theta^{}_{23}$ from $45^\circ$. It turns out that a normal
hierarchy of the light neutrino mass spectrum is favored in this
seesaw scenario.
\end{abstract}


\newpage

\framebox{\Large\bf 1} ~
Thanks to the enormous progress made in
solar, atmospheric and terrestrial neutrino experiments
\cite{SNO,SK,KM,K2K}, now we have got very robust evidence for the
existence of neutrino oscillations, a new window to physics beyond
the standard model. While the atmospheric neutrino deficit still
points toward a maximal mixing between the tau and muon neutrinos
($\theta^{}_{23} \simeq 45^\circ$), the solar neutrino anomaly
favors a large but non-maximal mixing between the electron and
muon neutrinos ($\theta^{}_{12} \simeq 33^\circ$). In addition,
the third neutrino mixing angle $\theta^{}_{13}$ is constrained to
the range $0^\circ \leq \theta^{}_{13} \leq 10^\circ$ and its
best-fit value is actually around zero \cite{Vissani}. These
results motivate a number of authors to consider the so-called
tri-bimaximal neutrino mixing pattern \cite{HPS}
\begin{equation}
V^{}_0 = \left ( \matrix{ 2/\sqrt{6} & 1/\sqrt{3} & 0 \cr
-1/\sqrt{6} & 1/\sqrt{3} & 1/\sqrt{2} \cr 1/\sqrt{6} & -1/\sqrt{3}
& 1/\sqrt{2} \cr} \right ) \; .
\end{equation}
It corresponds to $\sin^2\theta^{}_{12}=1/3$, $\sin^2
\theta^{}_{23} = 1/2$ and $\sin^2 \theta_{13}=0$. Such a special
neutrino mixing pattern is most likely to arise from an underlying
flavor symmetry (e.g., the discrete non-Abelian symmetry $A_4$
\cite{A4}). The spontaneous or explicit symmetry breaking is in
general unavoidable, because a flavor symmetry itself cannot
reproduce the observed lepton mass spectra and lepton mixing
angles simultaneously \cite{Grimus}.

Given the successful seesaw mechanism \cite{SS}, which naturally
explains the smallness of three (left-handed) neutrino masses, it
will be interesting to anticipate that $V^{}_0$ just results from
diagonalizing the effective neutrino mass matrix $M^{}_\nu$ in the
flavor basis where the charged-lepton mass matrix $M^{}_l$ is
diagonal and real (positive):
\begin{equation}
M^{}_\nu \; =\; M^{}_{\rm D} M^{-1}_{\rm R} M^T_{\rm D} \; = \;
V^{}_0 \overline{M}^{}_\nu V^T_0 \; ,
\end{equation}
where $M^{}_{\rm R}$ is symmetric and denotes the heavy
(right-handed) Majorana neutrino mass matrix, and
$\overline{M}^{}_\nu \equiv {\rm Diag}\{m^{}_1, m^{}_2, m^{}_3 \}$
with $m^{}_i$ (for $i=1,2,3$) being three light Majorana neutrino
masses. Then the tri-bimaximal neutrino mixing can be achieved in
three possible ways in such a seesaw model:
\begin{itemize}
\item      Taking $M^{}_{\rm D}$ to be diagonalized by $V^{}_0$
and $M^{}_{\rm R} = M^{}_0 {\bf I}$ with $M^{}_0$ being a common
mass scale and $\bf I$ being the identity matrix;

\item      Taking $M^{}_{\rm D}= m^{}_0 {\bf I}$ with $m^{}_0$
being a common Dirac fermion mass scale and $M^{}_{\rm R}$ to be
diagonalized by $V^{}_0$;

\item      Taking very simple but non-trivial textures of
$M^{}_{\rm D}$ and $M^{}_{\rm R}$ such that $V^{}_0$ results from
diagonalizing both of them.
\end{itemize}
Among these three possibilities, the first one is the simplest and
most interesting from a phenomenological point of view,  which will
be clearer later. The mass degeneracy of three heavy Majorana
neutrinos, which corresponds to the $S(3)$ flavor symmetry, has
actually been speculated in some interesting seesaw models
\cite{Yanagida,Zhou,XR,Joaquim2} so as to generate large neutrino
mixing angles. Although the well-motivated $V^{}_0$ is fairly
compatible with all today's neutrino oscillation data, a tiny
departure of the realistic neutrino mixing pattern $V$ from $V^{}_0$
remains allowed. In particular, possible deviation of
$\theta^{}_{23}$ from $45^\circ$ and non-vanishing $\theta^{}_{13}$
will be an important issue to probe at the upcoming neutrino
experiments. Thus it is worthwhile to consider the deviation of $V$
from $V^{}_0$ and to examine its possible origins \cite{rodej}. In
this paper, we aim to establish a direct relationship between the
breaking of heavy Majorana neutrino $S(3)$ symmetry and the
deviation of $V$ from $V^{}_0$. A similar idea has recently been
proposed in Ref. \cite{kk} to investigate the small correction,
induced by the mass splitting of heavy Majorana neutrinos, to the
well-known bi-maximal neutrino mixing pattern \cite{BI} in a simple
seesaw scenario.

We are going to begin with the first conjecture made above and
propose a simple scenario, in which the mass splitting between the
first generation and the other two generations of heavy Majorana
neutrinos is responsible for a departure of the solar neutrino
mixing angle $\theta^{}_{12}$ from its initial value $35.3^\circ$
given by the tri-bimaximal neutrino mixing pattern $V^{}_0$.
Furthermore, we shall show that the slight breaking of the mass
degeneracy between the second and third generations of heavy
Majorana neutrinos gives rise to a small mixing angle
$\theta^{}_{13}$ and a tiny deviation of the atmospheric neutrino
mixing angle $\theta^{}_{23}$ from $45^\circ$. These results imply
that our seesaw model has much more room to fit the present and
future neutrino oscillation data, and it provides some useful
hints towards further model building to obtain a dynamical picture
of neutrino mass generation and lepton flavor mixing.

The remaining part of this paper is organized as follows. Section
2 is devoted to some analytical calculations of how the breaking
of the mass degeneracy among heavy Majorana neutrinos is related
to the deviation of lepton flavor mixing from the tri-bimaximal
mixing pattern $V^{}_0$. A numerical illustration of the typical
parameter space is given in section 3. Some further discussions
and conclusion are presented in section 4.

\vspace{0.5cm}

\framebox{\Large\bf 2} ~
Let us conjecture that $M^{}_{\rm R} =
M^{}_0 {\bf I}$ holds in the symmetry limit and $M^{}_{\rm D}$ can
be diagonalized by the transformation $V^\dagger_0 M^{}_{\rm D}
V^*_0 = {\rm Diag} \{ x, y, z\}$ with $x$, $y$ and $z$ being real
and positive. Then Eq. (2) holds automatically. Note that it is
possible to decompose the tri-bimaximal neutrino mixing matrix
$V^{}_0$ into a product of two rotation matrices
$R^{}_{23}(\theta^{0}_{23})$ and $R^{}_{12}(\theta^{0}_{12})$;
i.e., $V^{}_0 = R^{}_{23}(\theta^{0}_{23}) \otimes
R^{}_{12}(\theta^{0}_{12})$, where
\begin{eqnarray}
R_{12}(\theta^{0}_{12}) & = & \left ( \matrix{ 2/\sqrt{6} &
1/\sqrt{3} & 0 \cr -1/\sqrt{3} & 2/\sqrt{6} & 0 \cr 0 & 0 & 1 \cr}
\right ) \; ,
\nonumber \\
R_{23}(\theta^{0}_{23}) & = & \left ( \matrix{ 1 & 0 & 0 \cr 0 &
1/\sqrt{2} & 1/\sqrt{2} \cr 0 & -1/\sqrt{2} & 1/\sqrt{2} \cr}
\right ) \; ,
\end{eqnarray}
with $\theta^{0}_{12} = \arctan (1/\sqrt{2}~) \approx 35.3^\circ$
and $\theta^{0}_{23} = 45^\circ$. In the following, we consider
two mass splitting scenarios for three heavy Majorana neutrinos
and calculate the corresponding light neutrino mass spectra and
flavor mixing angles.

\subsection{Scenario with $M^{}_1 \neq M^{}_2 = M^{}_3$}

In this case, we define the mass splitting parameter
$\delta^{}_{12} \equiv (M^{}_2 - M^{}_1 )/M^{}_2$ and assume
$|\delta^{}_{12}| \ll 1$. The effective neutrino mass matrix
$M^{}_\nu$ turns out to be
\begin{eqnarray}
M^{}_\nu & = & M^{}_{\rm D} M^{-1}_{\rm R} M^T_{\rm D}
\nonumber \\
& = & V^{}_0 \left ( \matrix{ x & 0 & 0 \cr 0 & y & 0 \cr 0 & 0 &
z \cr} \right ) \; V^T_0 \left ( \matrix{ M^{-1}_1 & 0 & 0 \cr 0 &
M^{-1}_2 & 0 \cr 0 & 0 & M^{-1}_2 \cr} \right ) \; V^{}_0 \left (
\matrix{ x & 0 & 0 \cr 0 & y & 0 \cr 0 & 0 & z \cr} \right ) V^T_0
\nonumber \\
& = & V^{}_0 \left ( \matrix{ x & 0 & 0 \cr 0 & y & 0 \cr 0 & 0 &
z \cr} \right ) R^T_{12} \left ( \matrix{ M^{-1}_1 & 0 & 0 \cr 0 &
M^{-1}_2 & 0 \cr 0 & 0 & M^{-1}_2 \cr} \right ) R^{}_{12} \left (
\matrix{ x & 0 & 0 \cr 0 & y & 0 \cr 0 & 0 & z \cr} \right ) V^T_0
\; \equiv \; V^{}_0 M'_\nu V^T_0 \; ,
\end{eqnarray}
where
\begin{equation}
M'_\nu \; = \; \frac{z^2}{3M^{}_1} \left ( \matrix{ \omega^2
\eta^2 \left (3-\delta^{}_{12} \right ) & \sqrt{2} ~\omega \eta^2
\delta^{}_{12} & 0 \cr \sqrt{2} ~\omega \eta^2 \delta^{}_{12} & ~~
\eta^2 \left (3-2\delta^{}_{12} \right ) ~~ & 0 \cr 0 & 0 & 3
\left (1-\delta^{}_{12} \right ) \cr} \right ) \;
\end{equation}
with $\omega \equiv x/y$ and $\eta \equiv y/z$. It is obvious that
$M^\prime_\nu$ can be diagonalized by the orthogonal
transformation $O^T_{12} M^\prime_\nu O^{}_{12} =
\overline{M}^{}_\nu$, in which $O^{}_{12}(\theta'_{12})$ denotes
the Euler rotation matrix analogous to $R^{}_{12}(\theta^0_{12})$,
and $\overline{M}^{}_\nu$ has been defined in Eq. (2). A
straightforward calculation yields
\begin{eqnarray}
m^{}_1 & = & \frac{z^2 \eta^2}{3M^{}_1} \left [\omega^2 \left
(3-\delta^{}_{12} \right )\cos^2\theta'_{12} + \left
(3-2\delta^{}_{12} \right ) \sin^2\theta'_{12} - \sqrt{2} ~\omega
\delta^{}_{12} \sin 2\theta'_{12} \right ] \; ,
\nonumber \\
m^{}_2 & = & \frac{z^2 \eta^2}{3M^{}_1} \left [\omega^2 \left
(3-\delta^{}_{12} \right )\sin^2\theta'_{12} + \left
(3-2\delta^{}_{12} \right )\cos^2\theta'_{12} + \sqrt{2}
~\omega\delta^{}_{12} \sin 2\theta'_{12} \right ] \; ,
\nonumber \\
m^{}_3 & = & \frac{z^2}{M^{}_1} \left (1-\delta^{}_{12} \right )
\; ;
\end{eqnarray}
and
\begin{equation}
\tan 2\theta'_{12} = \frac{2\sqrt{2} ~\omega \delta^{}_{12}}{3
\left (1 - \omega^2 \right ) + \left (\omega^2 -2 \right
)\delta^{}_{12}} \; .
\end{equation}
Then we obtain the lepton flavor mixing matrix
\begin{equation}
V \; = \; V^{}_0 O^{}_{12} \; = \; R^{}_{23}(\theta^{0}_{23})
\otimes R^{}_{12}(\theta^{}_{12}) \; ,
\end{equation}
where $\theta^{}_{12} = \theta^{0}_{12} + \theta'_{12}$. One can
see that the original mixing angle $\theta^{0}_{12}$ gets modified
as a consequence of the mass splitting between the first
generation and the other two generations of heavy Majorana
neutrinos. In the limit $\delta^{}_{12} \rightarrow 0$,
$\theta^{}_{12} \rightarrow \theta^{0}_{12}$ is therefore
restored. Because of $\theta^{}_{23} = 45^\circ$ and
$\theta^{}_{13} =0^\circ$, the deviation of $V$ from $V^{}_0$ is
fairly slight.

\subsection{Scenario with $M^{}_1 \neq M^{}_2 \neq M^{}_3$}

In this more general case, the mass degeneracy between the second-
and third-generation heavy Majorana neutrinos is lifted too. The
corresponding mass splitting parameter is defined as
$\delta^{}_{23} \equiv (M^{}_3 - M^{}_2 )/M^{}_3$. Of course,
$|\delta^{}_{23}| \ll 1$ is required to hold, because it is
expected to be responsible for the tiny deviation of
$\theta^{}_{23}$ from $45^\circ$ and for a small value of
$\theta_{13}$. The effective neutrino mass matrix $M^{}_\nu$ then
reads
\begin{eqnarray}
M^{}_\nu & = & M^{}_{\rm D} M^{-1}_{\rm R} M^T_{\rm D}
\nonumber \\
& = & V^{}_0 \left ( \matrix{ x & 0 & 0 \cr 0 & y & 0 \cr 0 & 0 &
z \cr} \right ) \; V^T_0 \left ( \matrix{ M^{-1}_1 & 0 & 0 \cr 0 &
M^{-1}_2 & 0 \cr 0 & 0 & M^{-1}_3 \cr} \right ) \; V^{}_0 \left (
\matrix{ x & 0 & 0 \cr 0 & y & 0 \cr 0 & 0 & z \cr} \right ) V^T_0
\; \equiv \; V^{}_0 M'_\nu V^T_0 \; ,
\end{eqnarray}
where
\begin{equation}
M^\prime_\nu \; = \; \frac{z^2}{6M^{}_1} \left ( \matrix{ \omega^2
\eta^2 \left (6-2\delta^{}_{12}-\delta^{}_{23} \right ) & \sqrt{2}
~\omega \eta^2 \left (2\delta^{}_{12}+\delta^{}_{23} \right ) &
-\sqrt{3} ~\omega \eta \delta^{}_{23} \cr \sqrt{2} ~\omega \eta^2
\left (2\delta^{}_{12}+\delta^{}_{23} \right ) & ~~ 2\eta^2 \left
(3-2\delta^{}_{12}-\delta^{}_{23} \right ) ~~ & \sqrt{6} ~\eta
\delta^{}_{23} \cr -\sqrt{3} ~\omega \eta \delta^{}_{23} &
\sqrt{6} ~\eta \delta^{}_{23} & 3 \left
(2-2\delta^{}_{12}-\delta^{}_{23} \right ) \cr} \right ) \; ,
\end{equation}
in which the terms of ${\cal O}(\delta^2_{12})$, ${\cal
O}(\delta^2_{23})$ and ${\cal O}(\delta^{}_{12} \delta^{}_{23})$
have safely been neglected. One can diagonalize $M^\prime_\nu$ by
using three Euler rotation transformations
$O^{}_{12}(\theta'_{12})$, $O^{}_{23}(\theta'_{23})$ and
$O^{}_{13}(\theta'_{13})$ in the following way: $O^T_{13} O^T_{23}
O^T_{12} M^\prime_\nu O^{}_{12} O^{}_{23}O^{}_{13} =
\overline{M}^{}_\nu$. After a lengthy but straightforward
calculation, we arrive at
\begin{eqnarray}
m^{}_1 & \simeq & \frac{z^2 \omega^2 \eta^2}{M^{}_1}
\left (1-\frac{1}{3}\delta^{}_{12}-\frac{1}{6}\delta^{}_{23} \right ) \; ,
\nonumber \\
m^{}_2 & \simeq & \frac{z^2 \eta^2}{M^{}_1} \left
(1-\frac{2}{3}\delta^{}_{12}-\frac{1}{3}\delta^{}_{23} \right ) \; ,
\nonumber \\
m^{}_3 & \simeq & \frac{z^2}{M^{}_1} \left
(1-\delta^{}_{12}-\frac{1}{2}\delta^{}_{23} \right ) \; ;
\end{eqnarray}
and
\begin{eqnarray}
\tan 2\theta'_{12} & \simeq & \frac{2\sqrt{2} ~\omega \left
(2\delta^{}_{12}+\delta^{}_{23} \right )}{6 \left (1-\omega^2
\right )+2 \left (\omega^2 - 2 \right )\delta^{}_{12} + \left
(\omega^2 - 2 \right ) \delta^{}_{23}} \; ,
\nonumber \\
\tan 2\theta'_{23} & \simeq & \frac{2\sqrt{6} ~\eta
\delta^{}_{23}}{6 \left (1-\eta^2 \right )+ 2 \left (2\eta^2 - 3
\right )\delta^{}_{12} + \left (2\eta^2 - 3 \right
)\delta^{}_{23}} \; ,
\nonumber \\
\tan 2\theta'_{13} & \simeq & \frac{-2\sqrt{3} ~\omega \eta
\delta^{}_{23}}{6 \left (1-\omega \eta \right ) + 2 \left
(\omega^2 \eta^2 - 3 \right )\delta^{}_{12} + \left (\omega^2
\eta^2 - 3 \right )\delta^{}_{23}} \;
\end{eqnarray}
in a good approximation. The neutrino mixing matrix is given by
\begin{eqnarray}
V & = & V_0 O^{}_{12} O^{}_{23} O^{}_{13}
\nonumber \\
& = & R^{}_{23}(\theta^{0}_{23}) \otimes R^{}_{12}(\theta^0_{12} +
\theta'_{12}) \otimes R^{}_{23}(\theta'_{23}) \otimes
R^{}_{13}(\theta'_{13}) \; .
\end{eqnarray}
We see that the original mixing angles $\theta^0_{12}$ and
$\theta^0_{23}$ are both modified, and the third mixing angle
$\theta^{}_{13}$ becomes non-vanishing.

It is interesting to take a look at the approximate analytical
dependence of $\theta^{}_{12}$, $\theta^{}_{23}$ and
$\theta^{}_{13}$ on $\delta^{}_{12}$ and $\delta^{}_{23}$. For
simplicity, let us assume that the magnitudes of $\omega$ and
$\eta$ are around $0.5$ or smaller. Then Eq. (12) can be
simplified to
\begin{eqnarray}
\tan 2\theta'_{12} & \simeq & \frac{\sqrt{2} ~\omega}{3 \left
(1-\omega^2 \right )} \left (2\delta^{}_{12}+\delta^{}_{23} \right
) \; ,
\nonumber \\
\tan 2\theta'_{23} & \simeq & \frac{\sqrt{6} ~\eta}{3 \left
(1-\eta^2 \right )}\delta^{}_{23} \; ,
\nonumber \\
\tan 2\theta'_{13} & \simeq & \frac{-\omega \eta}{\sqrt{3} \left
(1-\omega \eta \right )} \delta^{}_{23} \; .
\end{eqnarray}
Combining Eq. (13) with Eq. (14), we obtain
\begin{eqnarray}
\theta^{}_{12} & \simeq & \theta^0_{12} + \frac{\sqrt{2}
~\omega}{6 \left (1-\omega^2 \right )} \left
(2\delta^{}_{12}+\delta^{}_{23} \right ) \; ,
\nonumber \\
\theta^{}_{23} & \simeq & \theta^0_{23} +
\frac{\eta \left (2+\omega \right )}{3} \delta^{}_{23} \; ,
\nonumber \\
\theta^{}_{13} & \simeq & \frac{\eta \left (1-\omega \right
)}{3\sqrt{2}} \delta^{}_{23} \; .
\end{eqnarray}
This approximate result indicates that the mass splitting
$\delta^{}_{23}$ is responsible for both the non-vanishing
$\theta^{}_{13}$ and the slight deviation of $\theta^{}_{23}$ from
its original value $\theta^0_{23} = 45^\circ$. In comparison, the
small deviation of $\theta^{}_{12}$ from its initial value
$\theta^0_{12} \approx 35.3^\circ$ depends on both
$\delta^{}_{12}$ and $\delta^{}_{23}$. In the limits
$\delta^{}_{12} \rightarrow 0$ and $\delta^{}_{23} \rightarrow 0$,
$V \rightarrow V^{}_0$ is immediately restored.

\vspace{0.5cm}

\framebox{\Large\bf 3} ~
Scenario A is of course regarded as the
special case of scenario B in the limit $\delta^{}_{23}
\rightarrow 0$. The latter totally involves six independent
parameters: three eigenvalues of $M^{}_{\rm D}$ and three
eigenvalues of $M^{}_{\rm R}$; or equivalently, $z$, $\omega$,
$\eta$, $M^{}_1$, $\delta^{}_{12}$ and $\delta^{}_{23}$.
Correspondingly, there are six observable parameters: three
neutrino masses and three lepton mixing angles. Our present
knowledge on the neutrino mass spectrum and the flavor mixing
pattern is inadequate to fully determine the free parameters of
our seesaw model, but some useful constraints on its parameter
space ought to be obtainable.

To the leading order in our above calculation, two neutrino
mass-squared differences are given by
\begin{eqnarray}
\Delta m^2_{21} & \equiv & m^2_{2} - m^2_{1} \simeq \frac{z^4
\eta^4}{M^2_1} \left (1-\omega^4 \right ) \; ,
\nonumber \\
\Delta m^2_{32} & \equiv & m^2_{3} - m^2_{2} \simeq
\frac{z^4}{M^2_1} \left (1-\eta^4 \right ) \; .
\end{eqnarray}
The solar and atmospheric neutrino oscillation data yield $\Delta
m^2_{21} = (7.2 \cdot\cdot\cdot 8.9) \times 10^{-5}\ {\rm eV}^2$
and $\left|\Delta m^2_{32}\right| = (1.7 \cdot\cdot\cdot 3.3
)\times 10^{-3}\ {\rm eV}^2$ at the $99\%$ confidence level
\cite{Vissani}. Then we arrive at $0 < \omega < 1$, but both $0 <
\eta < 1$ and $\eta > 1$ are allowed. If $\eta > 1$ is assumed,
$\omega$ must be close to $1$ in order to assure $\Delta
m^2_{21}/|\Delta m^2_{32}| \sim {\cal O}(10^{-2})$ to hold. Note
that $\theta'_{12}$ serves as a perturbation to $\theta^{0}_{12}$
in our model, hence its reasonable magnitude is $\leq 5^\circ$.
Note also that $\omega \sim 1$ is undesirable, because it may lead
to a rather large value of $\theta'_{12}$ unless $\delta^{}_{12}$
and $\delta^{}_{23}$ are negligibly small. These arguments
motivate us to choose $0 < \eta < 1$ instead of $\eta > 1$; in
other words, the light neutrino mass spectrum favors a normal
hierarchy in our scenario.

In the standard parametrization of $V$, a global analysis of
current neutrino oscillation data yields $30^\circ \leq
\theta_{12} \leq 38^\circ$, $36^\circ \leq \theta_{23} \leq
54^\circ$ and $\theta_{13} \leq 10^\circ$ \cite{Vissani} at the
$99\%$ confidence level. One can see that the tri-bimaximal
neutrino mixing pattern $V^{}_0$ is actually in good agreement
with these data. Thus the perturbations $\theta'_{12}$,
$\theta'_{23}$ and $\theta'_{13}$ in scenario B of our seesaw
model should be well constrained.

Now we proceed to scan the whole parameter space relevant for two
neutrino mass-squared differences and three neutrino mixing angles
by using Eqs. (11), (12) and (13). Our numerical results are shown
in Figs. 1 and 2. Some comments are in order.

(1) The mass-squared differences $\Delta m^2_{21}$ and $\Delta
m^2_{32}$ are mainly determined by the parameters $\omega$, $\eta$
and $z^2/M^{}_1$. Fig. 1 shows that the allowed ranges of $\eta$
and $z^2/M^{}_1$ are very restrictive. In contrast, $\omega$ is
essentially unrestricted, although the region of $\omega
\rightarrow 1$ is apparently disfavored. Note that $m^{}_3 \approx
z^2/M^{}_1 \approx 0.05$ eV holds in the leading-order
approximation, as a straightforward consequence of the normal
neutrino mass hierarchy. Typically taking $\omega \approx 0.5$ and
$\eta \approx 0.45$, we obtain $m^{}_2 \approx 0.01$ eV and
$m^{}_1 \approx 0.0025$ eV. Given $z \sim 175$ GeV, a mass scale
close to the electroweak scale or the top-quark mass, it turns out
that $M^{}_1 \sim 6 \times 10^{14}$ GeV, which is just the typical
seesaw scale.

(2) Because both $\delta^{}_{12}$ and $\delta^{}_{23}$ serve as
small perturbations, we have taken $|\delta^{}_{12}| \leq 0.2$ and
$|\delta^{}_{23}| \leq 0.2$ in our numerical calculations. The
signs of these two parameters are crucial to control the
departures of $\theta^{}_{12}$ and $\theta^{}_{23}$ from
$\theta^{0}_{12}$ and $\theta^{0}_{23}$. Note that
$\theta^{}_{13}$ has been arranged to lie in the first quadrant
even for $\delta^{}_{23} < 0$, as one can see in Fig. 2. It is
likely to obtain $\theta^{}_{13} \geq 1^\circ$, provided
$\delta^{}_{23} \geq 0.1$ is taken. Note also that a small
departure of $\theta^{}_{12}$ from $\theta^{0}_{12}$ can be
achieved from non-vanishing $\delta^{}_{23}$ in the limit
$\theta^{}_{12} \rightarrow 0$, as shown in Eq. (15). In other
words, the mass splitting $\delta^{}_{23}$ can simultaneously
affect $\theta^{}_{12}$, $\theta^{}_{23}$ and $\theta^{}_{13}$,
while the mass splitting $\delta^{}_{12}$ mainly affects
$\theta^{}_{12}$ in our seesaw model.

We remark that the uncertainties of the values on $\sin^2\theta_{23}$ and
$\sin^2\theta_{12}$ are expected to reduce to about $20\%$ at $3\sigma$
in future experiments \cite{future}, which would constrain our parameters
$\delta_{23}$ and $\delta_{13}$ further up to $13\sim 18 \%$.
As for the mixing angle $\theta^{}_{13}$, we point out that the reasonable
magnitude of $\theta^{}_{13}$ in this seesaw scenario ($\theta^{}_{13} < 2^\circ$
for $|\delta^{}_{23}| \leq 0.2$) remains too small to be measured
in the near future.
Indeed, the sensitivity of a few currently-proposed reactor neutrino
oscillation experiments to $\theta^{}_{13}$ is at the level of
$\theta^{}_{13} \sim 3^\circ$ or $\sin^2 2\theta^{}_{13} \sim
0.01$ \cite{Wang}. In contrast, our scenario allows the mixing
angles $\theta^{}_{12}$ and $\theta^{}_{23}$ to deviate from their
original (tri-bimaximal) values up to $5^\circ$ for
$|\delta^{}_{12}| \sim |\delta^{}_{23}| \sim 0.2$. Such remarkable
deviations are expected to be observable in the upcoming
long-baseline neutrino oscillation experiments \cite{LBL}. Thus it
is possible to test the validity of our seesaw model even before
the neutrino factory era.

\vspace{0.5cm}

\framebox{\Large\bf 4} ~
We have proposed a very simple seesaw
model based on two phenomenological assumptions: (a) the masses of
three heavy Majorana neutrinos are exactly degenerate; and (b)
three light Majorana neutrinos have the tri-bimaximal mixing
pattern $V^{}_0$ in this limit. A small mass splitting between the
first generation and the other two generations of heavy Majorana
neutrinos is found to be responsible for the departure of the
solar neutrino mixing angle $\theta^{}_{12}$ from its initial
value set by $V^{}_0$. It is also shown that a small mass
splitting between the second and third generations of heavy
Majorana neutrinos gives rise to a non-vanishing mixing angle
$\theta^{}_{13}$ and a slight departure of the atmospheric
neutrino mixing angle $\theta^{}_{23}$ from $45^\circ$. In this
seesaw scenario, the mass spectrum of three light Majorana
neutrinos has a normal hierarchy, leading to a tiny effective mass
of the neutrinoless double-beta decay (of ${\cal O}(10^{-3})$ eV
or smaller).

It is worth remarking that radiative corrections to three neutrino
mixing angles, which can be evaluated by using the
renormalization-group equations from the seesaw scale to the
electroweak scale, are expected to be very small in our model. On
the one hand, the small mass splitting between any two generations
of three heavy Majorana neutrinos implies that the seesaw
threshold effects \cite{Threshold} are negligible. On the other
hand, the normal mass hierarchy of three light Majorana neutrinos
assures that the renormalization-group effect on three mixing
angles is insignificant when they run down to the electroweak
scale. These arguments are valid for both the tri-bimaximal
neutrino mixing pattern $V^{}_0$ and its modified form $V$ in the
framework of the standard model or its minimal supersymmetric
extension, unless $\tan\beta$ is extremely large \cite{Luo}.

Note that CP violation is not taken into account in our work. It is
well known that the exact mass degeneracy of three heavy Majorana
neutrinos forbids CP violation in their lepton-number-violating
decays \cite{Branco}, hence there is no thermal leptogenesis
\cite{LEP} in this case. If that mass degeneracy is lifted and
non-trivial CP-violating phases are introduced into $M^{}_{\rm D}$
and (or) $M^{}_{\rm R}$\cite{Joaquim1}, then it is possible to
achieve successful leptogenesis at the seesaw scale. It is in turn
likely to generate large CP violation in the neutrino mixing matrix
$V$ via the seesaw relation. While such ideas are certainly
interesting, they should be realized in a simple and suggestive way.
We shall explore the possibilities to combine our present seesaw
model with CP violation elsewhere.

Finally, we emphasize that the tri-bimaximal neutrino mixing
pattern is just a typical example in our seesaw scenario. One may
consider to embed other interesting (constant) patterns of lepton
flavor mixing, such as those proposed in Ref. \cite{Giunti}, into
our model in a similar way. In this sense, our phenomenological
scenario {\it does} provide a simple, useful and flexible
framework to understand the observed features of lepton flavor
mixing. It might also help provide some enlightening hints towards
further model building, in order to obtain a dynamical picture of
neutrino mass generation and leptonic CP violation.

\acknowledgments{SKK is supported in part by BK21 program of the
Ministry of Education in Korea and in part by KOSEF Grant No.
R01-2003-000-10229-0. The research of ZZX and SZ is supported in
part by the National Natural Science Foundation of China.}

\newpage

\begin{figure}[t]
\vspace{-2cm}
\epsfig{file=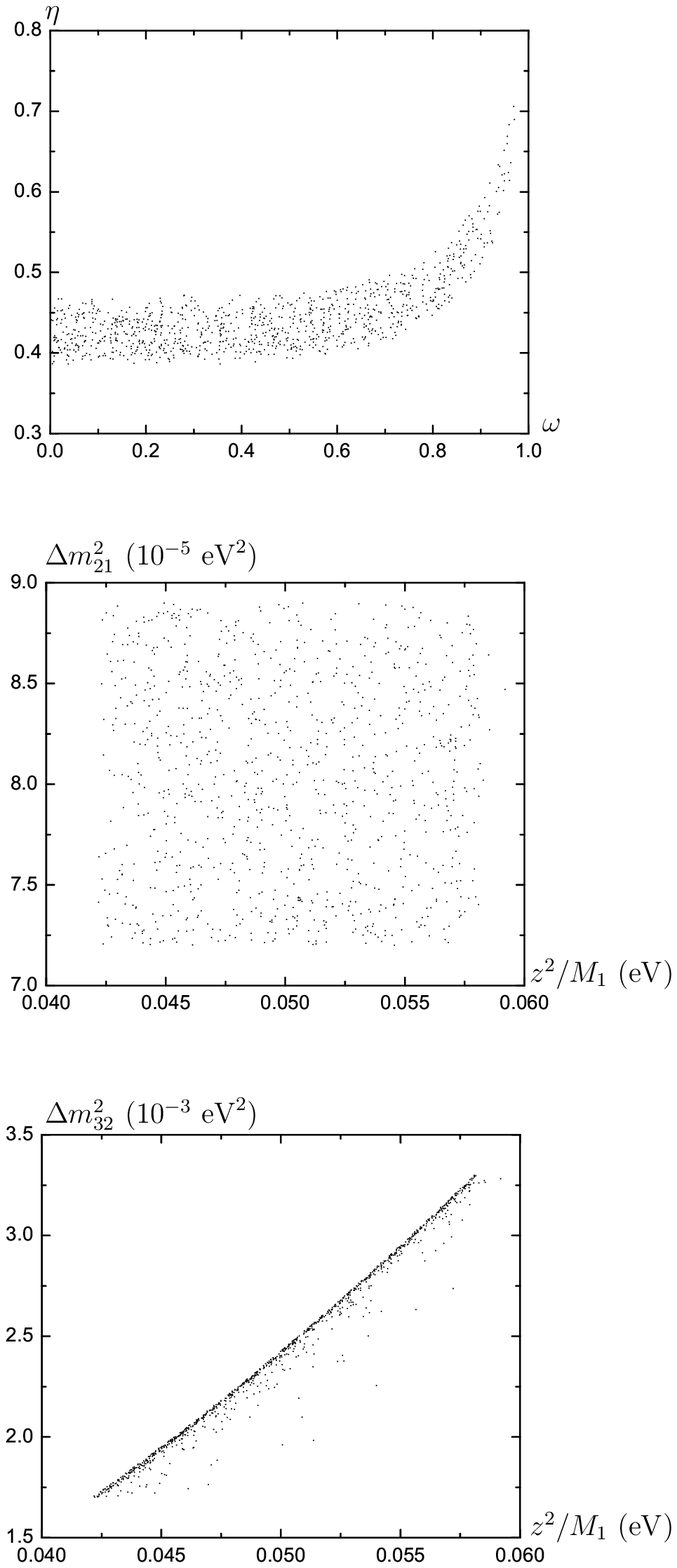,bbllx=0cm,bblly=19cm,bburx=5cm,bbury=29cm,%
width=4cm,height=9cm,angle=0,clip=0}  \vspace{14.2cm}
\caption{Allowed parameter space of ($\omega$, $\eta$),
($z^2/M_1$, $\Delta m^2_{21}$) and ($z^2/M_1$, $\Delta
m^2_{32}$).}
\end{figure}

\begin{figure}[t]
\vspace{-2cm}
\epsfig{file=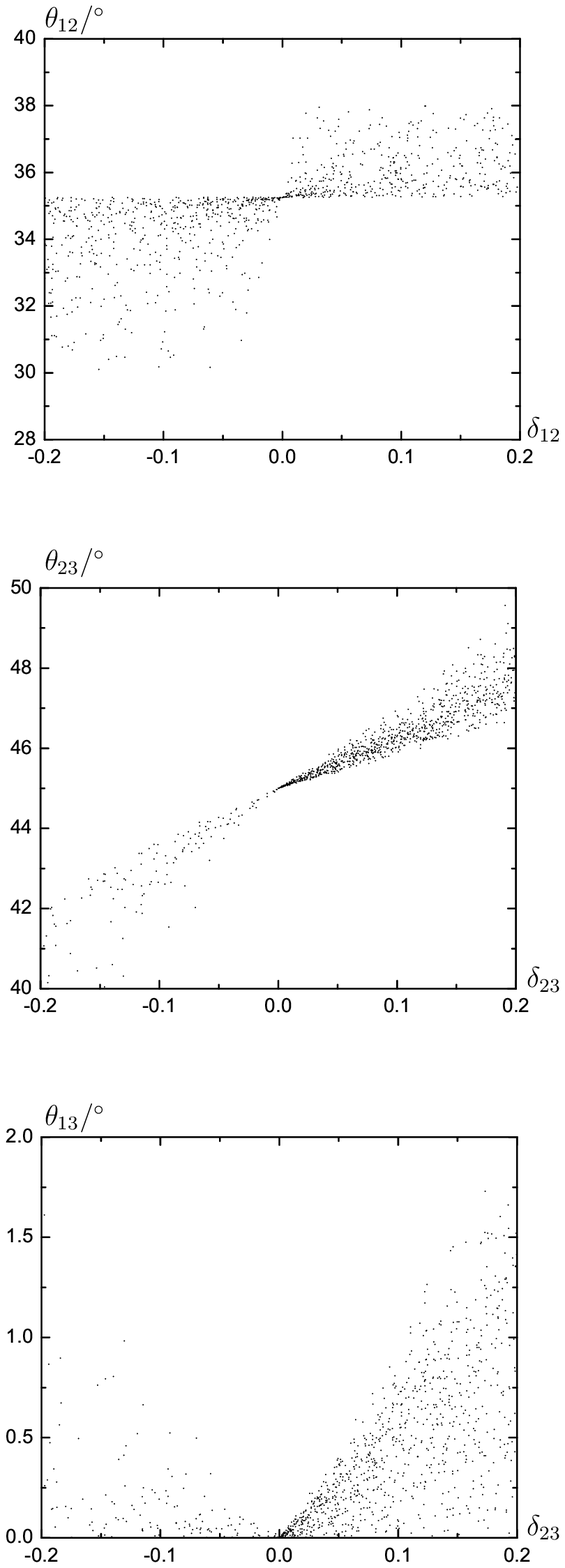,bbllx=0cm,bblly=16cm,bburx=5cm,bbury=26cm,%
width=4cm,height=9cm,angle=0,clip=0}  \vspace{12cm}
\caption{Allowed parameter space of ($\theta^{}_{12}$,
$\delta^{}_{12}$), ($\theta^{}_{23}$, $\delta^{}_{23}$) and
($\theta^{}_{13}$, $\delta^{}_{23}$).}
\end{figure}

\end{document}